\documentclass[useAMS,usenatbib]{mn2e}
\usepackage{graphicx}
\usepackage{amssymb,amsmath}
\usepackage{lscape}
\usepackage{url}
\title[Determination of Ceres mass based on the most gravitationally efficient close encounters]{Determination of Ceres mass based on the most gravitationally efficient close encounters}

\author[Andjelka B. Kova\v cevi\'c]{Andjelka B. Kova\v cevi\'c $^{1}$\thanks{E-mail:
andjelka@matf.bg.ac.rs (ABK)} \\
$^{1}$Department of Astronomy, Faculty of Mathematics, University of Belgrade, Studentski trg 16, 11000 Belgrade, Serbia\\
and Isaac Newton Institute of Chile, Yugoslavia Branch\\}

\begin{document}

\date{Accepted 2011 September 28.  Received 2011 September 28; in original form 2011
January 21}

\pagerange{\pageref{firstpage}--\pageref{lastpage}} \pubyear{2011}

\maketitle

\label{firstpage}

\begin{abstract}
Here is presented  recalculated value of the mass of Ceres, based on explicit tracking of its gravitational influence on orbits evolution of 21 selected asteroids during their mutual close encounters (CE). It was applied a new modified method (MM) for mass determination, based on the connecting of pre-encounter observations to the orbit determined from post-encounter ones. The calculated weighted mean value of Ceres mass, based on modified method, is $(4.54\pm0.07)\,10^{-10}M_{\odot}$ while standard procedure (SM) provided result of $(4.70\pm0.04)\,10^{-10}M_{\odot}$.
We found that correlation between individual estimated masses based on modified and standard method is $0.78$, which confirms reliability of using modified method.
\end{abstract}


\section{Introduction}
The main asteroid belt (MAB) is a living relic, with developing collisional
and dynamical evolution slowly erasing traces of planet accretion processes.
Until now it is well known that better knowledge of the masses of the larger asteroids helps to improve dynamical model of our Solar system.

Besides this, the beginning of XXI century gives rise to new possibilities of importance of asteroid mass determination.
We know that there are scaling laws that control collisional evolution both during and after planetary accretion (especially the end of the XX  and beginning of the XXI century have seen an developing  interest in
asteroid mass determinations, particularly in the context of the growing awareness
of the threat of asteroidal impact upon Earth.)
Fragmentation of asteroids is consistent, as suggested by \cite{Pio53}, with the observed size distribution which is equivalent to the differential mass distribution.

 As for mass distribution of our Solar system,  one can see by using data from Allan's Astrophysical Quantities \citep{Co00}, for planets, satellites and brighter asteroids (diameters of this asteroids and some satellites were transformed into masses assuming density of $2 \rm{g/cm^{3}}$), that present very rough mass function of Solar system bodies (see Fig. 1) is incomplete below $10^{-9} M_{\odot}$ (roughly Pluto mass).
 Keeping in mind that the asteroids are not all composed of the same material
and are not  compact to a similar degree, we will just illustrate how would look like possible construction of a low-mass extension of this function from faint asteroids. All bodies are assumed
to be spherical and are set to the same density. We
set the bulk density of each body here to be  $2 \, \rm{g/{cm^3}}$. This
value is a kind of a compromise between our simple assumptions, need for easier calculation and the measured
densities of several different groups: the average bulk densities
of several multi-km C-type asteroids ($1.3\,\rm{ g/ {cm^3}}$), the
average bulk density of several multi-km S-type asteroids
($2.7\, \rm{g/ {cm^3}}$), and the grain densities of several different
meteorite classes that may be more representative of the
bulk densities of sub-km asteroids ($2.2\, \rm{g/{ cm^3}}$ for CI meteorites,
$2.7\, \rm{g/{cm^3}}$ for CM meteorites, $3.5\,\rm {g/{cm^3}}$ for CV
meteorites; $3.5-3.8\,\rm{g/ {cm^3}}$ for ordinary chondrites) \citet{BR02}.

 \citet{I01} gave a diameter distribution function for 13000 asteroids based on the Sloan Digital Sky Survey. They set values of asteroids  albedo between 0.14 (for S type) and of 0.04 (for C type). The inferred diameter ($D$) distribution had power law form $N(D) \,{\rm{d}}D\propto D^{-\gamma} \,{\rm{d}}D$ (${\rm{d}}N/{\rm{d}}D$ is asteroid differential size distribution) where values of parameter $\gamma$ have value of 4 for asteroids larger than 5 km, while for smaller asteroids it is  about 2.3. This is in good agreement with  later studies performed by \citet{T05}. If we take into account simple relationship between the mass and diameter of an asteroid as well as mentioned value of  $2 \, \rm{g/{cm^3}}$ for average asteroid density than we can obtain mass function which is of the same form as it is distribution function, $\phi(M) \,{\rm{d}}M\propto M^{-\gamma^\prime} \,{\rm{d}}M$ with an exception that power law index, ${-\gamma^\prime}$ is -2 for asteroids larger than 5 km  and -1.1 for smaller bodies. This distribution covers mass range from $10^{-20} M_{\odot}$ (asteroids with $D\sim 0.3 \, \rm{km}$) to $10^{-13} M_{\odot}$ (asteroids with $D\sim 100 \, \rm{km}$). It can be seen that distribution has a bump (knee) near $D~5\, \rm{km}$.
Of course, because of the lack of adequate data, the Mass Function amplitude
for small bodies, is likely underestimated.

Also, we used  available {\it IRAS\/}  data for asteroids and their average density $2 \,\rm{g/{cm^3}}$ in order to obtain mass distribution. We used method introduced by  \citet{Kr01} and obtained mass distribution with respect to the asteroid radii, shown in Fig.2. We also obtained value $M(0)$ when $r\to 0$. The result of fitting  process is the following $M(r)=11.75-0.298r^{0.5}$, $M(r)$ are given in solar masses (units $10^{-10} M_{\odot}$) and r in kilometers. This result is comparable with result obtained by using the mapping of the crater size frequency
distribution of each planetary surface
to a corresponding projectile size-frequency distribution
(by application of an adequate
scaling law, see \cite{HN11}).
This estimated mass of MAB is about $4\%$ of the mass of the Moon. The largest object in MAB is Ceres, which contains about $32\%$ (or  $1/3$)  of the MAB total mass. The IAU currently recognizes a few dwarf planets in our Solar System, but only two of these bodies, Ceres and Pluto, have been observed in enough detail to demonstrate that they fit the definition.

So, if we take into account that Ceres causes very significant perturbations on almost all bodies in MAB and that it fits well into the definition of dwarf planet, one can say that the knowledge of its mass is of great dynamical and taxonomic importance.
Until now,  the
greatest number of asteroid mass determinations is based on the perturbations exerted on asteroids during their close approaches with massive MAB body. However, such masses have quite low
accuracies, but are still the best available for most asteroids. From statistical point of view, the use of as many perturbed asteroids as possible is crucial for reliable estimation of the asteroid mass. We therefore used the already known most suitable close approaches (20) and also performed a search for newly ones, which led us to use 24 close encounters for the determination of Ceres mass. (Here we will note that The MBA is a gravitationally chaotic system \citep{BC08} in such a way that asteroid mass determination could be compared to isolation of a very weak signal from almost white noise background. For example, in the case of our selected past close encounters with Ceres, we found that signal to noise ratio was $\log(24/40000)=-3.221$ ("Signal-to-noise ratio" is sometimes used informally to refer to the ratio of useful information to false or irrelevant data), which means that noise is greater than signal. So, detecting  of the gravitationally most effective close encounters in MBA for asteroid mass determination is a serious task based on available observations.
The main goal of our work is new mass determination  of Ceres based on the new method. This new method assumes mapping of 7 dimensional vectorial space of pre(post)encounter osculating elements  of perturbed body ($\{E_1, E_2, \dots , E_6\}$) and  perturbing mass ($m$) into 1-dimensional vectorial space of RMS of post(pre)encounter part of orbit:

 \begin{equation}
(\{E_1, E_2, \dots , E_6\}, m)\longmapsto RMS\,, \end{equation}

This mapping is based on numerical procedure where actually the correct mass of perturbing body will give the best fit of post(pre)encounter observations of perturbed body with pre (post) encounter part of perturbed orbit. The innovation is that perturbing mass is determined independently from corrections of perturbed osculating elements. Also, we use some new  close encounters for  Ceres mass determinations. In such a way we also demonstrate usefulness of close encounters with high numbered bodies.


\begin{figure*}
   \includegraphics[width=8.8cm]{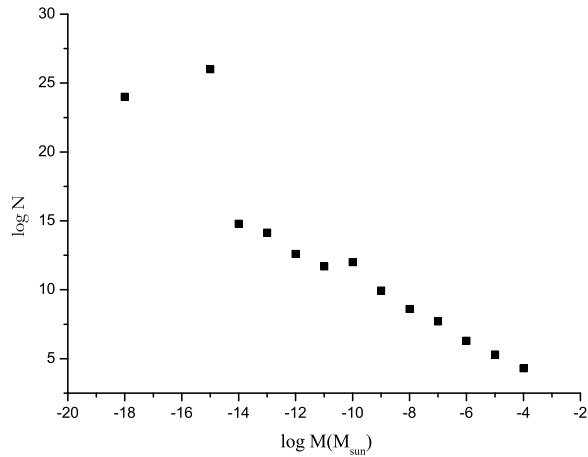}
 \caption{ Mass distribution function for solar planets, satellites and small bodies given in solar mass (units $10^{-10} M_{\odot}$) and r are given in kilometers . Squares correspond to differential number of bodies. Bodies which diameter is smaller than 1km were not included. The power law index for this fit is -0.93.}
\label{distrib}
\end{figure*}


\begin{figure*}

\includegraphics[width=8.8cm]{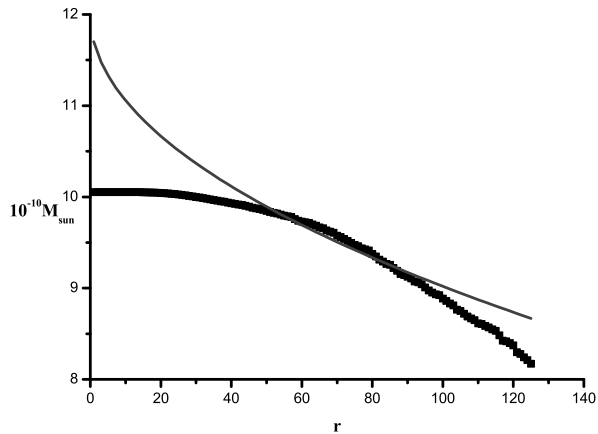}
\caption{ Mass distribution function for asteroids, based on {\it IRAS\/}  data (bold black dots). $M(r)$ (units $10^{-10} M_{\odot}$) is total mass of bodies which diameter is greater than $r$ (in kilometers). The fitted curve is presented as continuous line.}
\label{distrib2}
\end{figure*}

\section{Historical Ceres mass determinations}
 Ceres was intensively observed by different techniques.  Due to  the
NASA space mission Dawn (\citet{Rs06}), which had been launched in 2007,
the mass of Ceres  will be quite well estimated in
the very near future. Namely, it should reach the vicinity of Ceres in February 2015.
Historical Ceres mass determinations are presented in Fig. 3. As it can be seen, the mass values distribution significantly fluctuate up until 2000, after 2000 fluctuations are stabilized between $(4.7-4.8)\,10^{-10}M_{\odot}$. Further, it could be seen significant decreasing in formal errors of results since 1997, perhaps induced by new observational techniques.
Also, we can see 3 year long gaps in Ceres mass determinations (the first between 1992 and 1995, and the second between 2001 and 2004). As it can be seen from the calculated parameters (given in Table 1) of distributions of Ceres mass values and corresponding formal errors, one can see that there is a significant variability among previously published mass determinations, which validates the need for additional study.


 So for the first time, here are presented redetermined, by means of two methods (standard and modified method), masses of Ceres based on the most gravitationally efficient CE. Ceres is suitable for this purpose due to its physical characteristics and the fact that its mass will be soon determined with high accuracy.

\begin{figure*}
\centering
\includegraphics[width=8.8cm]{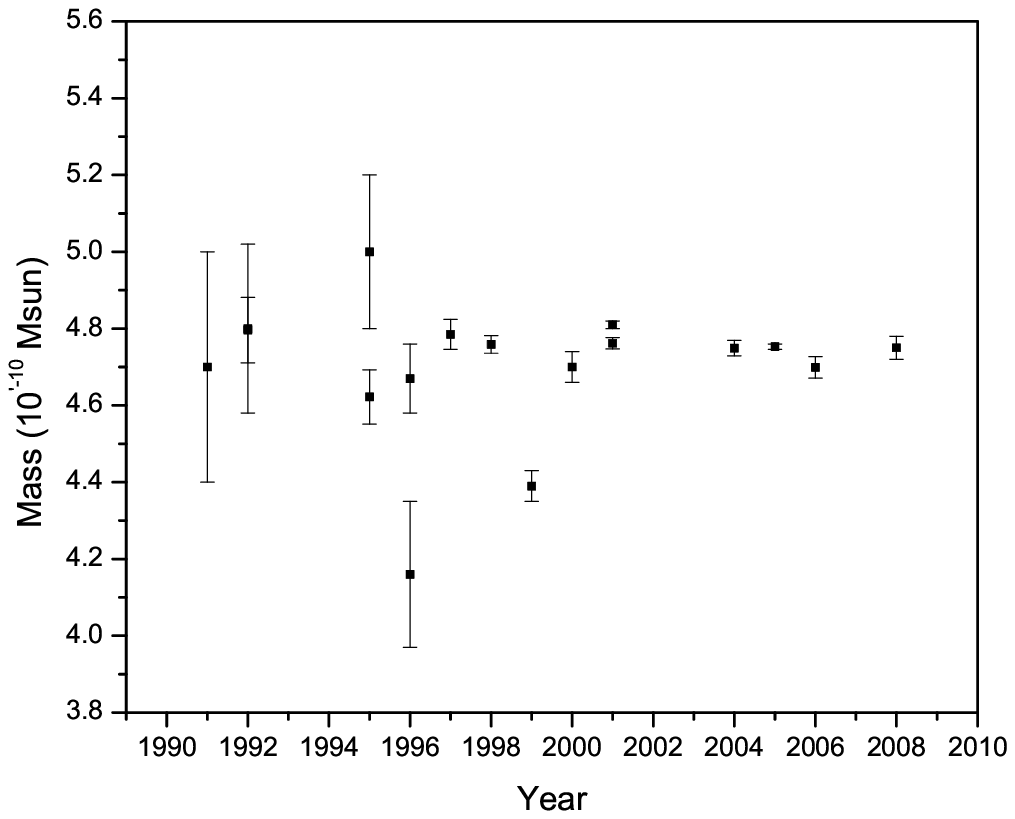}
\caption{Historical time curve of Ceres mass determinations. Goffin 1991,  	
Sitarski and Todorovic-Juchniewicz 1992, Williams 1992, Sitarski and Todorovic-Juchniewicz 1995, Viateau and Rapaport 1995, Carpino and Kne\v zevi\' c 1996, Kuzmanoski 1996, Viateau and Rapaport 1997, Viateau and Rapaport 1998, Hilton 1999, Michalak 2000, Standish 2001, Pitjeva 2001, Pitjeva 2004, Pitjeva 2005, Konopliv et al. 2006, Baer and Chesley 2008.}
\label{history}
\end{figure*}

   \begin{table*}
   \centering

 \caption{Descriptive statistics of historical values of Ceres mass. Columns designation: N is total number of values, SD is standard deviation, V is variance, S is skewness, K is Kurtosis, GM is geometric mean, GSD is geometric standard deviation, Min is minimum value, Med is median value, Max is maximum value.}
 \label{historystat}
\begin{tabular}{lllllllllll}

      \hline
    &   N& SD  & V   &S &K &GM&GSD&Min&Med&Max  \\
 \hline

Mass Values&	17&	0.18&	0.0336&	-1.797&	4.63&	4.697&	1.041&	4.16&	4.75&	5\\
Formal errors&	17&	0.089&	0.0079	&1.370	&0.806&	0.048&	3.053&	0.007&	0.04&	0.3\\
       \hline
    \end{tabular}
   \end{table*}

 \section{Procedure of mass determination}

We compiled list of historical CE with Ceres, already used for its mass determination, which gravitational efficiency is larger than
$10^{\prime\prime}$ in right ascension and we added four high numbered (which were not used previously) perturbed bodies: Cheng, Lyzenga, Tasman and 2001QW120 (Table 2). This additional four high numbered bodies we selected according to a complex, combined procedure based on multistep selection and numerical integration. First, using a simple geometrical consideration, we find all the pairs with Ceres whose osculating orbits, are close enough to enable a real approach of these bodies. After that, by means of two body dynamics, we check, for each pair, the occurrence of such approach within  the given time span. At the end, the two orbits are numerically integrated  within  dynamical model in order to determine more precisely parameters of the CE (epoch, distance, relative velocity, deflection angle, etc.) and make definite conclusion.
Computation of the minimum distances between the orbits of the two asteroids involves the determination of the true anomalies of the relative nodes which serve as initial values and an iterative derivation of the corresponding true anomalies of the points where the orbits are closest to each other. We used only pairs with minimum distance between orbits of the two bodies less than 0.05AU.
As we mentioned, in order to find which pair of asteroids on nearby orbits can have CE in a given time, we apply the simple two body approximation and first compute the mean anomaly of the point corresponding to the orbital proximity of one of the asteroids and associated initial epoch. Then we calculate mean anomaly of the other body in the pair corresponding to the same initial time instant and to the subsequent time instances obtained by adding  the time interval equal to the first body's revolution period up to 50 years backward. Finally, we look at the given time span for all occurrences of the other body in the vicinity of the proximity point (within the range of $\pm 1^{\circ}$).
After completing these steps for  Ceres, we found about 40000 pairs satisfying our criteria. The previous steps had provided only the unperturbed data, so as the next step, we have to include the effects of perturbations. We performed a numerical integration for all pairs, covering in each case the interval  from the common osculation epoch of the orbital elements to an epoch which is 10 days before  possible CE, then we integrated both orbits for the next 20 days with step of 0.5 and 0.1 days. As a final step we calculated for selected asteroids gravitational perturbations induced by Ceres mass (the differences in  right ascension) and choose those objects which right ascension is larger than $10^{\prime\prime}$. In such a way, we found all asteroids which are well known as perturbed by Ceres mass but also we found those new four asteroids.
Geometrical and kinematical parameters, as well as expected gravitational effects (see Table 2), revealed the potential high efficiency of these close approaches. Actually, parameters $\rho$,$V_{r}$ and $\theta$ reflect potential gravitational efficiency of CE and further investigation is needed (which will be explained further). The most obvious criterion for selection of perturbed minor planets is the value of minimum distance, $\rho$, between it and perturbing body: the less value is the better. \citet{Hiel96} supposed that $\rho$ should be no greater than 0.05 AU. We have 18 bodies which had minimum distance no greater than mentioned value. As it was shown in \cite{KC01}, the number of perturbed bodies can be substantially enlarged as a result of consideration of bodies moving in the vicinity of commensurability of mean motions with perturbing minor planets. In this case, one can take into consideration not only close ($<0.05\,\rm{AU}$) encounters, but moderate (which occurs at distances $<0.1\, \rm{AU}$) as well.
Although the using of a less strict condition  could be justifiable as the approaches of planets may occur over and over due to commensurability, \cite{Ho89}  pointed out fruitfulness of such approaches.

We have 6 such approaches ($<0.1\, \rm{AU}$). Actually, many of encounters given in Table 2 are simply the closest of many encounters. Among them, it is well known that Ceres and Pallas are  in near commensurability. A resonance requires that the gravitational perturbation of one body upon the
other to keep the latter librating with respect to some critical variable. This is
not the case with Ceres and Pallas. In fact, actual asteroid-asteroid resonances are
quite rare, see for example \cite{Ch00}. Almost every 4.6 years these two largest bodies mutually approach. Between 1820 and 1839 (see Fig 7) all close encounters were within $0.25\, \rm{AU}$ (the closest encounter, $0.18\, \rm{AU}$ was in 1820). The given encounter was among the closest recently. However, each preceding and subsequent encounter will have an effect on the perturbed asteroid (Pallas). In the case of  $1:1$ resonance in the mean motions, the changes increase due to the cumulative effect.  However,  for a near resonance (as it is in the case of close approach between Ceres and Pallas) the cumulative effect may be positive  or
negative. For example, \cite{Hi97} shows that the total perturbation
of 15 Eunomia on 1313 Berna is positive, but for 1284 Latvia although individual
perturbations may be quite large the total perturbation is small. Also note that
Latvia and Berna have eccentricities and inclinations similar to that of Eunomia,
unlike Ceres and Pallas. Furthermore, the difference between the mean semi-major
axes of Eunomia and Latvia is only half the difference between those of Ceres and
Pallas.
Also, the changes of their orbits constantly increase due to perturbing effects by major planets.

On account of this fact, very important is application of  the procedure of the numerical integration in order to calculate differences in right ascension and declination, which is described above (in dynamical model were included 9 largest bodies in MAB and all major planets and Pluto). In such a way we accounted gravitational influence of multiple encounters and give calculated values in Table 3 ($\Delta{\alpha}$ and $\Delta{\delta}$).

Also, as a measure of potential efficiency of a given encounter one can apply the formula for an asymptotic angular deflection, given by \cite{O76} and later applied by \citet{Hiel96}, as a potential  criterion of selection of pairs (in Table 2, parameter $\theta$  is change of trajectory of perturbed body due to gravitational influence). Under the condition that the mass of the larger body is much greater than perturbed (this is satisfied in the case of CE of Ceres), this term is the same as the equation for Rutherford scattering. Main effect of an asteroid-asteroid encounter is a velocity change, which causes a cumulative orbital longitude excursion. Actually, relative velocity gives insight into orbital confinement  in the fastest diverging orbital elements (longitude of node and pericenter) only, disregarding other orbital elements. And this parameter is also sustained in the term of deflection angle. We can see that our close encounters have $\theta$ less or equal to  $1.0\,^{\prime\prime}$. Since the encounters are usually nearly coplanar the predominant changes to the orbit of perturbed asteroid are in semi major axis. The change in this orbital element can be seen in the cumulative difference in position over time. So we give calculated maximum differences in right ascension and declination as last two columns in Table 3. So, the deflection angle and difference in right ascension and declination reveled that more moderate encounters (which occur at distances less than $0.1\, \rm{AU}$) could be efficient for Ceres mass determination. Again,  we will emphasize that deflection angle reflects potential possibility of using close encounter for mass determination, so we further calculated gravitational perturbations caused by Ceres mass (and major planets and Pluto) on motion of perturbing bodies ($\Delta{\alpha}$ and $\Delta{\delta}$ given in Table 6).

We also presented the case of (34755) 2001 QW120 as an illustration of the use of highly numbered asteroids. This asteroid is observed since 1950. Actually there are two observations from 1950: the first was taken  17.03.1950 at 11h 17m 42.33s and the second was taken 17.03.1950 at 11h 17m 44.02s. We tried to determine the mass of Ceres, based on MF, using both observations as a preencounter orbit and obtained the value of $3.94\,10^{-10}M_{\odot}$.  Then we performed further experiment, and tried to determine Ceres mass which will connect calculated postencounter orbit and only one preencounter observation taken in 11h 17m 42.33s. We obtained the value of $4.86\,10^{-10}M_{\odot}$ for Ceres mass, and this value of Ceres mass gives for the  O-C for preencounter observation: $0.784587$ arcsec and  $-0.070147$ arcsec in right ascension and declination,  respectively. However, the same experiment with preencounter observation taken in 11h 17m 44.02s, provided value for  Ceres mass $4.31\,10^{-10}M_{\odot}$ and this value of Ceres mass gives for the O-C of preencounter observation: $-0.0005$ arcsec and   $-0.019877$ arcsec  in right ascension and declination,  respectively. So, we used about 92 observations of which is only one preencounter (which gives the best O-C). Consequently, one can say that  it is used to illustrate such cases where there is only one preencounter observation.

 The past CE data are visualized in the time-minimum distance-relative velocity space in Fig. 5. The past CE are concentrated between 1960 and 2000, and occurred, mostly, at the distances within the range 0.05 and 0.3 AU. It could be explained by limitations of accuracy of observational astrometric technique used.

For the purpose of comparison, we calculated a list of  15 (Table 3) asteroids which will have close approach with Ceres within the distances less than 0.02 AU.
It could be seen that $40\%$ of found perturbed bodies have deflection angle (due to influence of Ceres mass) greater than $1^{\prime\prime}$. Having in mind that average relative velocity of CE is $5 \, {\rm {km/sec}}$, it could be seen that majority of future CE have relative velocities bellow this value (Fig. 5). Actually, it is not surprising, since the magnitude of the angle of the ballistic deflection is inversely proportional to the square of the relative speed. Thus, the cases with highest deflections would be clearly more likely to involve encounters at low relative speed. Only in the case of CE (1,3687) this value is moderately larger (Table 4).
Changes of relative velocities $\Delta V_{2}$ are large in the case of the majority of CE, which implies that significant gravitational effects (caused by Ceres mass) could be expected.
The test which we use to compare distributions of CE of Ceres are based on Kolmogorov-Smirnov test (\cite{Kolm33},\cite{Smir39}).
Its nonparametric and unbinned nature makes it well suitable to the sparsely distributed (about 20
events spread over multiple decades and parameter dimensions) CE population, for the difference of the binned $\chi^{2}$ test.
In its one-dimensional form, the Kolmogorov-Smirnov statistic is relatively simple to implement. First, one forms a cumulative
histogram from each data set, and normalizes both to one. These are estimates of the Cumulative Distribution
Functions (CDFs) from which the data sets were drawn, and are sometimes referred to as 'Empirical Distribution
Functions' (EDFs). The Kolmogorov-Smirnov statistic,
 is the maximum difference between the two EDFs. Because the size of this difference does not change under
reparameterization of the x axis (i.e., arbitrary increase or decrease of the separation between data points), the Kolmogorov-Smirnov
test is nonparametric. For the standard
one-dimensional Kolmogorov-Smirnov test, the confidence level is given by \cite{Pel02}.
This standard Kolmogorov-Smirnov test is sensitive only to variation in the CDFs of the two populations from which the data sets
are drawn. As long as the EDFs of the two data sets are similar, the test will report that the starting hypothesis is true,
even if the number of counts in the two sets is very different.

For example, we consider the case where the statistical properties of a sample $\{x_0, x_1, \dots , x_{n-1}\}$  obtained from a repeated experiment using a continuous random
$\mathbb{P}$
variable, should be compared to a given distribution function $F_{X}(x)={\mathbb{P}}(X\leq x)$.

The basic idea of the Kolmogorov-Smirnov test is to compare the distribution
function to the empirical sample distribution function
\begin{equation}
F_{\hat{X}}(x)=\mathbb{P}_{\hat{X}}(\hat{X}\leq x)=\frac{1}{n}\sum^{n}_{i=1}I(\hat{X_{i}}\leq x)
\end{equation}
Note that $F_{\hat{X}}$ is piecewise constant with jumps of size $1/n$ at the positions
$x_i$ (assuming that each data point is contained uniquely in the sample). It counts the proportion of the sample points below level x.
For any fixed points $x\in\mathbb{R}$ the law of large numbers implies that
\begin{equation}\begin{split}
F_{\hat{X}}(x)=\frac{1}{n}\sum^{n}_{i=1}I(\hat{X_{i}}\leq x)\rightarrow \\
\rightarrow \mathbb{E}I((\hat{X_{i}}\leq x)=\mathbb{P}(X\leq x)=F_{X}(x)
\end{split}\end{equation}
i.e. the proportion of the sample in the set $(\infty,x]$ approximates the probability of this set.
It could be shown from here that this approximation holds uniformly over all $x\in\mathbb{R}$

 \begin{equation}
d_{max}\equiv \sup_x |F_{X}(x)-F_{\hat{X}}(x)| \rightarrow 0 \end{equation}
i.e. the largest difference between $F_{\hat{X}} $ and $F_{X}$ goes to 0 in probability. The key observation in the Kolmogorov-Smirnov test is that the distribution of this supremum does not depend on the 'unknown' distribution $\mathbb{P}$ of the sample, if $\mathbb{P}$ is continuous distribution.
Since the sample distribution function changes only at the sample points,
one has to perform the comparison just before and just after the jumps. Thus, Eq. 4  is equivalent to
\begin{equation}
d_{max}\equiv \max_{x_i} (|F_{X}(x_i)-{1/n}-F_{\hat{X}}(x_i)|, |F_{X}(x_i)-F_{\hat{X}}(x_i)|) \end{equation}

The p-value, i.e. the probability of a value of $d_{max}$ as measured ($d^{measured}_{max}$ )
 is approximately
given by (see \cite{Pel02} and references therein):
\begin{equation}
P(d_{max}\geq d^{measured}_{max})=Q_{KS}(\langle \sqrt{n}+0.12+0.11/{\sqrt{n}}\rangle d^{measured}_{max})
\end{equation}
This approximation is already quite good for $n\geq 8$. Here, the following auxiliary
probability function is used:
\begin{equation}
Q_{KS}(\lambda)=2\sum^{\infty}_{i=1}(-1)^{i+1}e^{-2i^2\lambda^2}
\end{equation}
where $Q_{KS}(0)=1$ and $Q_{KS}(\infty)=0$.
Kolmogorov-Smirnov normality test (see Table 4) shows that at the 0.05 level of confidence, a significant number of the past and future CE are significantly drawn from a normally distributed population as a function of time, mutual distances  and relative velocities. To understand the test, it is helpful to see, (as an example) in Fig 8, how looks like an empirical cumulative distribution plot and  theoretical normal distribution for mutual distances of future close encounters with Ceres.

\begin{table*}
 \caption{Kinematical characteristics of past CE with (1) Ceres.}
 \label{kcp}
\resizebox{8cm}{!} {
\begin{tabular}{lllll}

      \hline
       Perturbed& Date  & $\rho$   &$V_{r}$ &$\theta$    \\

       asteroid& & [\,$\mathrm {AU}$]  & [\,$\mathrm {km\, s^{-1}}$] & [$^{\prime\prime}$] \\

            \hline

      (2) Pallas & 16.05.1825&    0.188& 12.61& 0.01\\

      (32) Pomona& 25.11.1975&    0.025&  4.75& 0.31\\

      (76) Freia&  05.08.1957&    0.212&  4.08& 0.05\\

      (91) Aegina&  13.09.1973&    0.033&  3.28& 0.49 \\

     (203) Pompeia&  22.08.1948&    0.016&  4.12& 0.63 \\

     (347) Pariana&  29.05.1943&    0.078&  1.48& 1.02\\

     (348) May&  02.09.1984&    0.046&  0.79& 6.07 \\

     (454) Mathesis&  23.11.1971&    0.021&  2.93& 0.97 \\

     (488) Kreussa&  17.07.1963&    0.282&  3.00& 0.07 \\

     (534) Nassovia& 24.12.1975&    0.023&  2.75& 1.00 \\

     (548) Kresida& 13.07.1982&    0.049&  2.95& 0.41 \\

     (621) Werdandi&  01.05.1962&    0.050&  3.04& 0.38 \\

     (741) Botolphia&  07.11.1940&    0.102&  1.36& 0.92 \\

     (792) Metcalfia&  25.07.1950&    0.013&  5.78& 0.40 \\

     (811) Nauheima&  09.10.1968&    0.030&  3.73& 0.42 \\

     (850) Altona& 22.02.1970&    0.026&  3.84& 0.45 \\

    (1642) Hill&25.11.1925&    0.012&  5.54& 0.47 \\

    (1847) Stobbe&07.09.1958&    0.094&  1.76& 0.60\\

    (2572) Annschnell&26.03.1971&    0.012&  4.79& 0.63 \\

    (3344) Modena&27.09.1980&    0.021&  2.39& 1.45 \\

    2051 Chang&  22.10.1943&    0.012&  4.48& 0.72 \\

    6010 Lyzenga& 29.04.1973&    0.011&  8.03& 0.24 \\

    6594 Tasman& 15.05.1982&    0.013&  6.71& 0.30 \\

    34755 2001QW120& 27.11.1967&    0.005&  8.35& 0.49\\

  \hline
       \end{tabular}
   }
   \end{table*}

\begin{table*}
 \caption{Kinematical characteristics of future CE with (1) Ceres.}
 \label{kc}
\resizebox{8cm}{!} {
\begin{tabular}{llllll}

      \hline
       Perturbed& Julian Date  & $\rho$   &$V_{r}$ &$\theta$ &$\Delta V_{2}$     \\

       asteroid& & [\,$\mathrm {AU}$]  & [\,$\mathrm {km\, s^{-1}}$] & [$^{\prime\prime}$]&$[{\rm{km/ sec}}\cdot 10^{-9}]$ \\

            \hline
  1393&  19.05.2022& 0.006914& 2.213& 5.43&  58302\\
   3687& 22.09.2060& 0.011792& 9.347&  0.18&   8092\\
     3795& 27.08.2059& 0.018447& 3.401&  0.86&  14217\\
   4286&18.07.2049& 0.012773& 3.740& 1.03&  18670\\
  4882&17.12.2046& 0.012384& 4.521&  0.73&  15931\\
  5955& 27.05.2045& 0.013210& 5.379&  0.48&  12552\\
   6093&29.12.2055& 0.017281& 5.088&  0.41&  10144\\
  6212&05.07.2049& 0.017693& 6.483&  0.25&   7776\\
  6813& 28.12.2014& 0.016464& 4.327&  0.60&  12519\\
  7407& 03.11.2028& 0.011996& 4.121&  0.90&  18042\\
  7426& 21.10.2028& 0.014137& 4.610&  0.61&  13686\\
  7855&10.02.2026& 0.006095& 5.442& 1.02&  26887\\
   8893& 13.06.2041& 0.013787& 7.617&  0.23&   8493\\
 10246& 24.11.2032& 0.015615& 5.201&  0.44&  10983\\
 11348& 15.05.2054& 0.014660& 3.938&  0.81&  15448\\
        \hline
    \end{tabular}
  }
   \end{table*}

\begin{table*}
 \caption{Kolmogorov-Smirnov test for future and past CE. The Kolmogorov-Smirnov statistic,
 is the maximal difference between the two Empirical Distribution Functions. Both, past and future sets of CE with Ceres are significantly drawn from normally distributed population, at the 0.05 level, as function of time, mutual distances and relative velocities.}
 \label{kcfpce}
\resizebox{8cm}{!} {
\begin{tabular}{llll}

      \hline
       CE& Parameter  & Statistic &Probability     \\

            \hline

Future&Time &0.15&0.95\\
      &$\rho$ & 0.18&0.70\\
       &$V_{r}$ & 0.21&	0.49\\
  Past&  Time&   0.21&	0.21\\
         &$\rho$ &   0.29&	0.03\\
         &$V_{r}$ &0.17 &	0.43\\

       \hline
    \end{tabular}
  }
   \end{table*}

   \begin{figure*}
\includegraphics[totalheight=5.3cm]{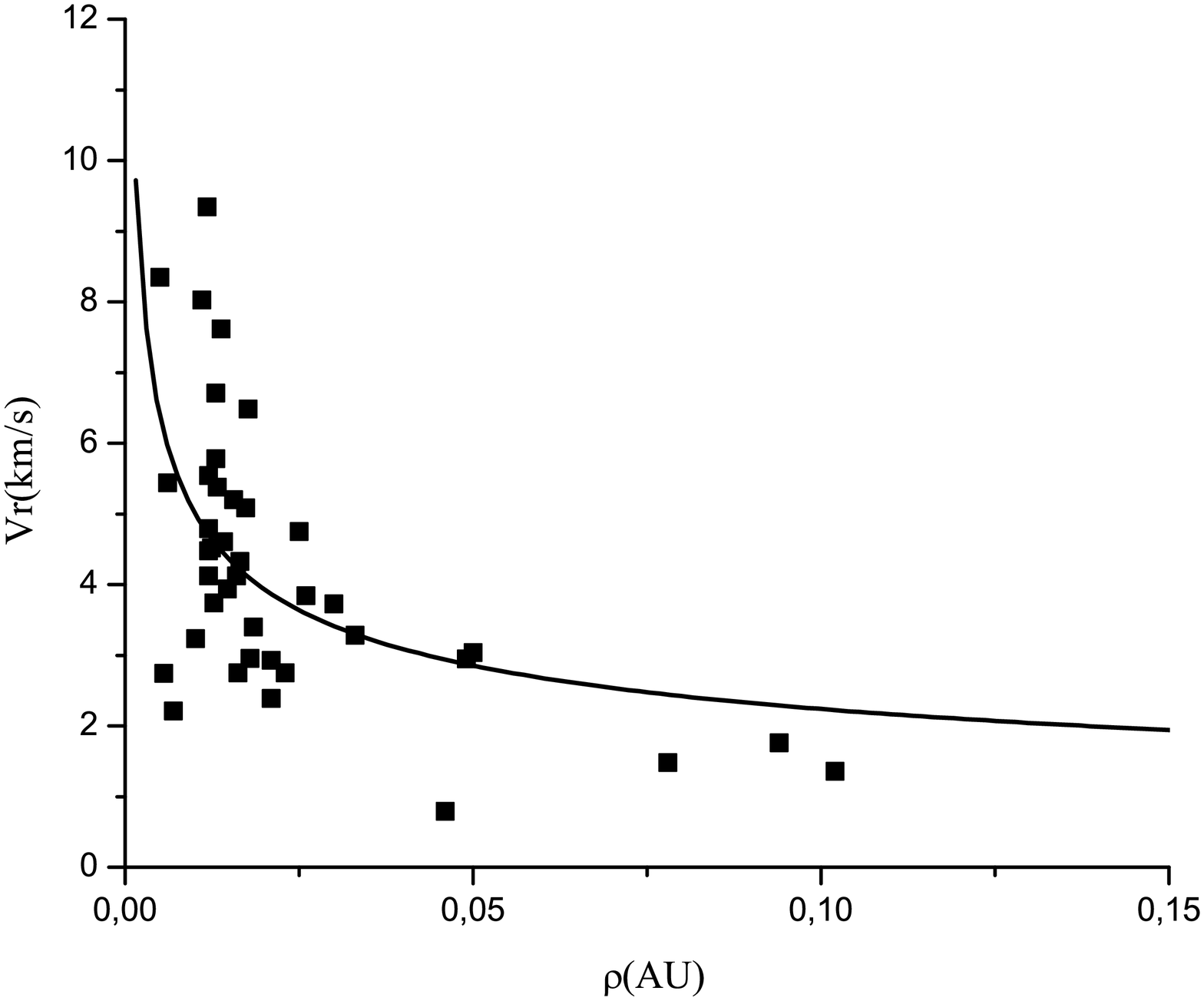}
\includegraphics[totalheight=5.3cm, origin=c]{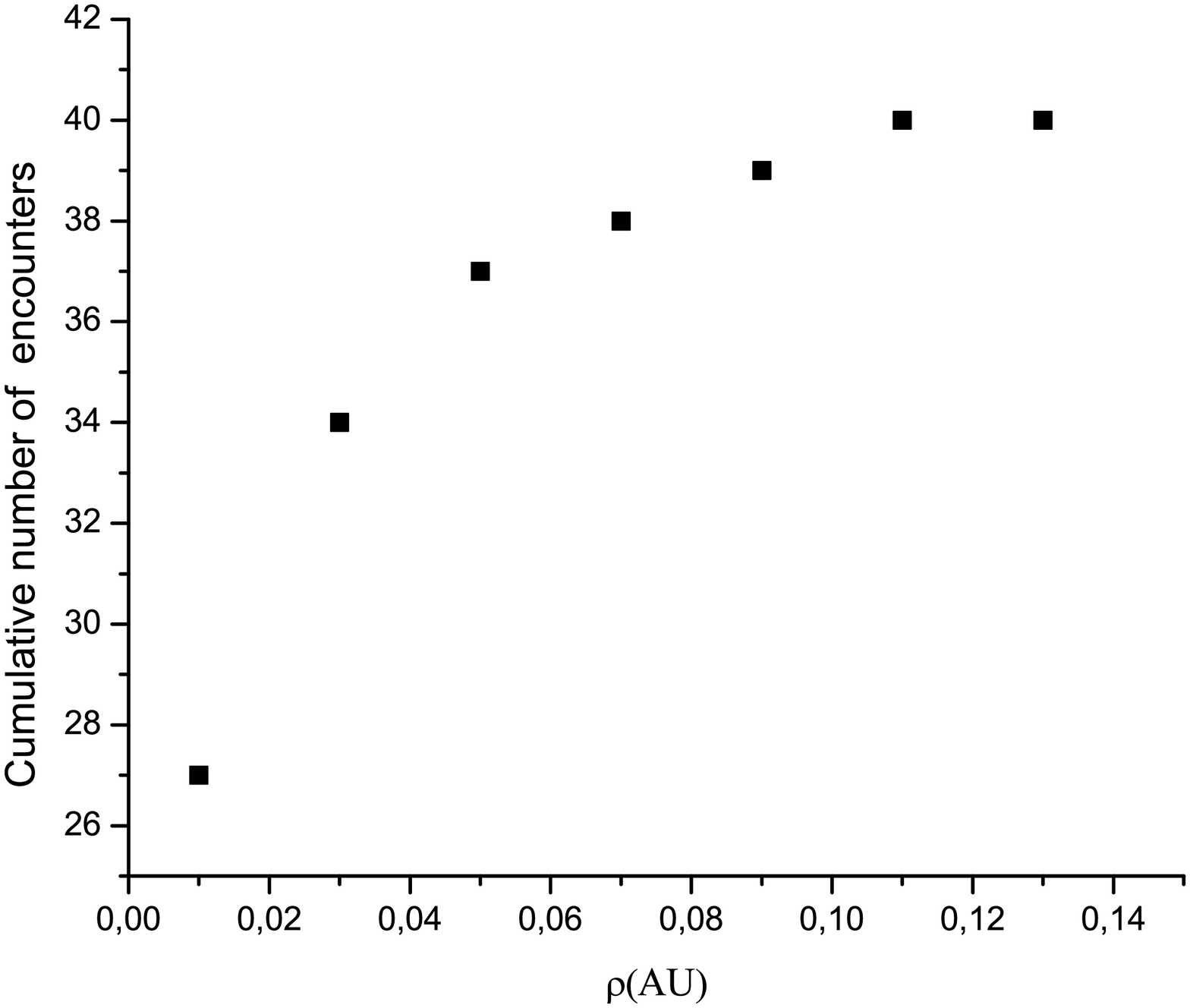}
\caption{Joined past and future CE with Ceres. Left panel: Solid curve is power function fitted with power index: $(-0.35\pm 0.01)$.
Right panel: Distribution of cumulative number of past and future CE with Ceres.} \label{vr}
\end{figure*}


\begin{figure*}
\centering
\includegraphics[width=8.8cm]{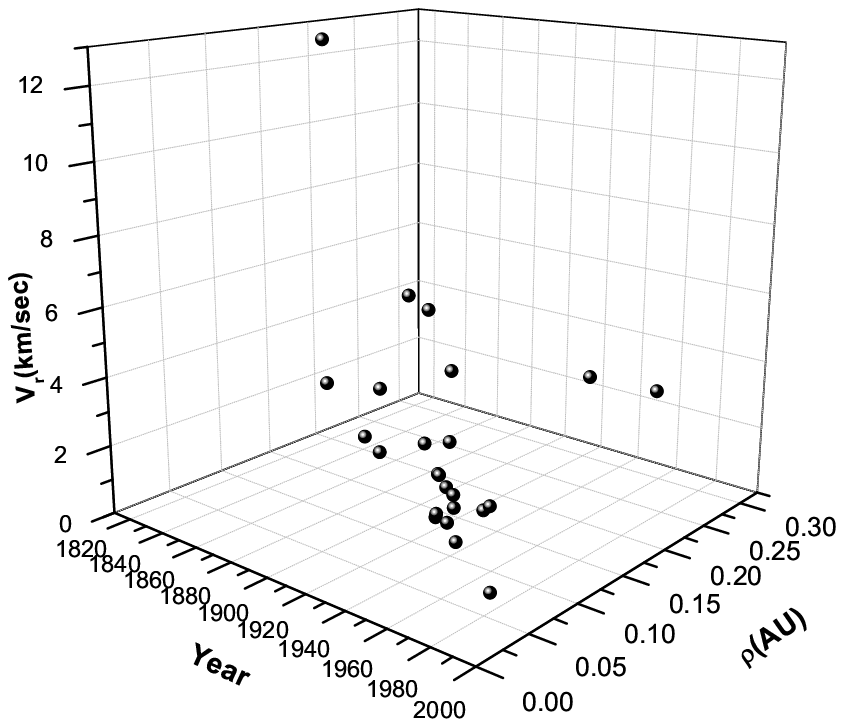}
\includegraphics[width=8.8cm]{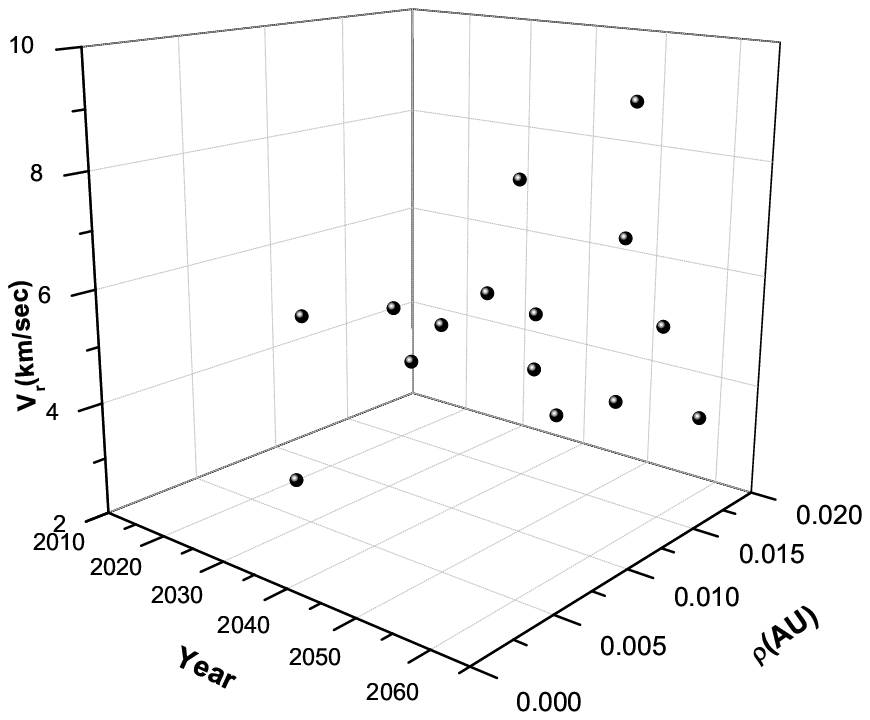}
\caption{Left panel: The set of selected past CE with Ceres depicted in time, mutual distance and relative velocity phase space. The distribution shows the dense region between 1960 and 2000. In this region  relative velocities are smaller than $5\, \rm{km/sec}$. The closest CE occurred in 1967 with 34755 at the distance nearly to the radius of Ceres Hill sphere (0.0015 AU). Right panel: The set of selected future CE with Ceres depicted in time, mutual distance and relative velocity phase space. The distribution is without dense regions. There are two CE, in 2022 and 2026, which occurred at the distances nearly to the Ceres Hill sphere radius. Majority of CE have relative velocities smaller than $6\, \rm{km/sec}$.}
\label{ceres2}
\end{figure*}

We joined past and future sets of CE which could be seen in Figure 4. The acceleration due to the gravity of Ceres mass is clearly manifested for distances smaller than 0.05 AU and their cumulative number is growing rapidly at the same range of mutual distances.
This is a manifestation of the fact that the magnitude of ballistic deflection is inversely proportional to both the relative velocity and the minimum distance.The mean value of relative velocity of joined sets of CE is $4.25 \, {\rm{km/s}}$. The Pearson correlation coefficient for the data on the right plot is $0.88\pm 0.01$ (the most distant point is discarded).
The mass determination of Ceres was performed by means of modified method (MM)  introduced by \cite{Ku96} which is fully developed and applied in \cite{Ko05, Ko06}, and the classical least-square method (SM), widely used by many authors. The 9 largest asteroids have been included in the dynamical model, as well as all major planets.
The essence of MM approach is the following: to separate pre and postencounter orbits of the perturbed asteroid. In the calculation of these two orbits it is not necessary to know the mass of perturbing asteroid, because its perturbing effects are practically negligible (zero value). These two orbits are separated by an impulsive change due to the close encounter and can be connected by properly accounting for gravitational effects of the perturbing body. If the pre and postencounter orbits are precisely determined, the correct mass of perturbing body will give the best fit of postencounter observations with the preencounter orbit and vice versa, the preencounter observations with the postencounter orbit. With such an approach, we first calculated pre and postencounter orbits. With the postencounter orbit we  fitted preencounter observations, setting the trial values for the mass of Ceres and computing O-C residuals in the RMS sense.

From algorithmical point of view, the MM procedure of asteroid mass
determination consists of:
(i) calculation of the orbit of the perturbed asteroid using
only postencounter  observations and
(in the first iteration) the adopted value of the perturbing mass (due to its influence during other close encounters at greater distances or in the case of temporary
resonances);
(ii) determination of the correction $\Delta m$ of the mass of the
perturbing asteroid by means of the classical least-square
method. This correction is the solution of the system
of linear equations based on the preencounter (or postencounter)
observations

\begin{equation}
\Delta m={(H^{T}H)}^{-1}{H}^{T}R\,,
\end{equation}

\noindent where H is the symmetric and positive definite matrix of the partial derivatives (because the corrections are small and the problem has been linearized)
of the coordinates (right ascension and declination) of the perturbed body with respect to the
perturbing mass m, and R is the matrix of (O-C) residuals,
in coordinates, of the perturbed body.
 Since the dimension of R is greater than the dimension of $\Delta m$ we will have least square solution for the mass correction. Note that term on right hand side of equation is the pseudo invers of the matrix H.
The differences between two methods may be divided into theoretical differences and differences in the input data.
Modified method separates perturbed orbit in two parts (pre and postencounter part), excluding
 the part nearby the CE \citep{Ku10}. In such a way, the mass of perturbing body is not dominant in the calculated mean motion of perturbed body.
The modified method corrects  the mass of the perturbing asteroid by means of the least square solution of the system
of linear equations based on the preencounter (or postencounter)
observations.

We also derived expression for formal error for asteroid mass determined by MM as

\begin{equation}
\sigma_{m}={{\sigma_{0}}\over{\sqrt{\sum_{i=1}^n{\partial
c_{i}\over\partial m}}}}\,, \end{equation}

where n is number of used pre or postencounter observations, $c_{i}$ are coordinates (right ascension and declination), while  $\sigma_{0}$ is given by term:
\begin{equation}
\sigma_{0}={\sqrt{(O-C)^2\over{{2n-1}}}}\,,
\end{equation}

We  can see that this expression includes intoself (not all) some  parameters (mentioned by \citet {Bo94}) of the semi-empirical metric which could parameterize error:  O-C positional residuals, coordinates and number of accurate observations (before and after encounter) of perturbed body.
In ideal case the length of pre and postencounter orbit would be almost equal, allowing the best fitting results. However
asymmetry in the length of pre and postencounter parts of an orbit and uneven observation distribution could affect calculation process.
In order to determine Ceres mass, we used asteroids listed in Table 2, which also have large enough observation covering. All test asteroids (except the last four bodies), were used by previous investigators and they are well known as strongly perturbed bodies by Ceres.
The last four bodies listed in Table 2 were found in a different way: combining traditional
approach and procedure introduced by \cite{Ku92}. The outcome of this procedure was the
list of dates of the CE of (1) Ceres
with suitable perturbed asteroids as well as the absolute value of the maximum difference in right ascension and declination between two trajectories of perturbed body: the first one takes into account perturbation of (1) Ceres, whereas the second does not.
If the difference was large (typically, larger than 10 or 15
arcsec in right ascension) and if the available observations covered long enough period before and after
the encounter, the perturbed asteroid was selected
as a good candidate for the mass determination.

Here we will describe the manner of data (from different observatories and
different eras) handling. How data are handled is critical because often the
signal-to-noise is quite low. For example, the closest approach of Pallas to Ceres
occurred in the early $19^{\rm{th}}$ century. At that time the uncertainty in single positions
was high, of the order of few arcseconds, and existed significant  errors in
star catalogues. Or early observations of a particular perturbed asteroid
may be significantly less accurate if taken decades before the more recent
observations. The quality of observations, particularly older observations, may vary
significantly from observatory to observatory (Table 6, columns designed with RMS).
For this purpose, we applied combination of criteria for estimating
the quality of the observations and discarding the less reliable
data. Firstly,  we discarded all observations which have AstDys residual rms greater than 5 arcsec.

In the calculation of the post-encounter orbits
of perturbed asteroids, the $3\sigma$ criterion for the selection of
observations was applied. All observations that had a residual
above $3\sigma$ in at least one of the coordinates ($\alpha$, $\delta$) were discarded.
However, having in mind that preencounter observations are often of poorer quality and sparse, the same criterion in the case of preencounter
orbit calculation gives very different results in the RMS sense
for different perturbed asteroids. Because of reasoning to use as much as possible useful preencounter
observations, in these cases we discarded all observations where
the discrepancy in each of the coordinates was larger than 2.5
arcsec. Finally, in the case of mass determination using the
standard method, we discarded observations that had a residual
larger than 2.5 arcsec in at least one of the coordinates. As it is
known  there are much more newly collected observations than
older ones. Also, the new observations are of higher accuracy.
In the orbit calculation, these facts result in a small standard
deviation of observations (see in Table 6 columns $RMS_{1}$ and $RMS_{2}$). Because of that, by using
the $3 \sigma$ criterion, many old observations, which are necessary
for mass determination, would be discarded.

\section{Results and discussion}

 In comparison with historical results (Fig. 3), masses that we obtained (Table 6) by using standard method are in good agreement  (except the case of close approach with (792) Metcalfia).
In  the cases of CE with (32) Pomona, and (1847) Stobbe, obtained Ceres masses are different than the adopted value ($4.76\,10^{-10} M_{\odot}$) in this paper (this value is slightly larger than Current Best Estimates for Ceres mass adopted by the IAU 2009 GA). Actually, these masses are out of $3\sigma$ distance from adopted value.
As we have two series of Ceres mass  values obtained by MM and SM, the main question is  could  they be summarized to some value. Especially, if we take into account that some values in our data sample might be known to be more variable (less precise) than other values. Since the results of individual calculations of Ceres mass are based on non-intersecting data sets, the values for the mass of Ceres supplied by each perturbed asteroid can be considered as uncorrelated and therefore can be combined into a unique value through a weighted average.(Speaking in term of analysis, it is possible to summarize series of Ceres mass values by weighted mean method, because we could make subsequences of calculated masses and also make subsequences of weights (calculated formal errors,  which have tendency to converge to zero)). Weighted mean values are calculated according to the well known formula:
 \begin{equation}
 \bar{x} = \frac{ \sum_{i=1}^n w_i x_i}{\sum_{i=1}^n w_i},
 \end{equation}
where $\{x_1, x_2, \dots , x_n\}$ are a set of calculated Ceres masses, while  $w_{i}=1/{\sigma_{i}}^2$ are weights (where $\sigma_{i}$ is formal error of corresponding Ceres mass determination). The
weighting scheme can greatly affect the adopted value, so knowledge of how the
weights were determined is critical. For
example, they might be the uncertainty in the individual weights, or
the size of the pre and postencounter residuals, or even equal weights. Having this in mind,  we will explain more carefully how the weights themselves were arrived.
Weighted functions are commonly used in statistics to compensate for the presence of bias. For example, for a quantity f measured multiple independent times $f_i$  with variance $\sigma^2_i$, the best estimate of the f is obtained by averaging all the measurements with weights $ w_i=\frac 1 {\sigma_i^2}$, and the resulting variance is smaller than each of the independent measurements ( $\sigma^2\sim1/\sum w_i$). Besides this, we will mention that maximum likelihood method weights the difference between fit and data using the same approach.
As we generated our data series from numerical experiments, there will be some error in the variance of each data point. Such errors may be underestimated due to not taking into account all sources of error in calculating the variance of each data (as it was mentioned earlier, in our formal error of Ceres mass are not included all parameters). So the formal error must be  calculated as
\begin{equation}
 \sigma_{\bar{x}}^2 = \frac{ 1 }{\sum_{i=1}^n 1/{\sigma_i}^2} \times \frac{1}{(n-1)} \sum_{i=1}^n \frac{ (x_i - \bar{x} )^2}{ \sigma_i^2 }.
 \end{equation}
 where term which follows $\times$ is the calibration coefficient, which is an estimate of how much the real dispersions of single determinations with respect to the average value are larger than the formal standard deviations of each individual mass determination.
 So, we calculated weighted mean Ceres mass without mentioned cases, and obtained $(4.70\pm0.04)\,10^{-10}M_{\odot}$ which is within  standard $3\sigma$ distance from adopted value of Ceres mass. Actually, as we said, we summarized our data series according to weighted mean method and obtained  meaningful result (close to adopted value). Also, we calculated the formal error of such algebraic operation.
\begin{table*}
   \centering
\caption{Descriptive statistics of calculated Ceres masses by standard (SM) and modified method (MM). Columns designation: N is total number of values, SD is standard deviation, V is variance, S is skewness, K is Kurtosis, GM is geometric mean, GSD is geometric standard deviation, Min is minimal value, Med is median value, Max is maximal value. Parameters of Kolmogorov-Smirnov normality test: St is statistics, Prob is probability. Level of confidence is 0.05}

 \label{descript}
\begin{tabular}{lllllllllllllll}

      \hline
    &   N& Mean&SD  & V &Sum  &S &K &GM&GSD&Min&Med&Max &St&Prob \\
 \hline

SM&21&	4.73&	0.41&	0.17&	99.91&	0.54&	1.20&	4.74&	1.09&	3.94&	4.74	 &5.81&0.12&1\\
err SM&21	&0.24&	0.26&	0.07&	4.87&	2.30&	5.63&	0.14&	2.62&	0.04&	0.16&	 1.1&0.23&0.17\\

MM&21&	4.63&	0.36&	0.13&	97.73&	0.13&	-1.17&	4.64&	1.08&	4.1	& 4.68&	5.22&0.12&0.97\\
err MM&21&	0.12&	0.12&	0.01&	2.4	&1.91	&4.00&	0.07&	3.14&	0.01	&0.08&	 0.49&0.23&0.19\\

       \hline
    \end{tabular}
   \end{table*}
As the next step of our investigations, we calculated Ceres masses based on modified (Table 6) method.

Weighted mean value of Ceres mass, obtained from modified method, is $(4.54\pm0.07)\,10^{-10}M_{\odot}$.
Some pairs had the secondary CE which occurred at the distances greater  than radius of gravitational sphere of Ceres (as it is in the cases of close approaches with Pallas and (1642) Hill).
Both, SM values and MM have mean, geometric mean and median values very close (Table 5). However, MM express negative kurtosis (shape of distribution has lower, wider peak around the mean). The skeweness of SM and MM values is positive, while in the case of MM it is closer to zero, which means that this distribution is more even. Kolmogorov-Smirnov normality test has shown (see Table 5 and Fig 9)  that a significant number of SM and MM values and their corresponding errors  at the 0.05 level of confidence are significantly drawn from normally distributed population.
For normally distributed data we should expect about $15\%$ of the data to lie more than 1 standard deviation below the mean (i.e., below $4.70-0.41=4.29$ in the case of standard method, and $4.54-0.36=4.18$ in the case of modified method), in fact there is only one value  (standard method values) and two values  in the case of modified method.
Error values of MM express closer mean, geometrical mean and median values than error values of SM results. Distributions of both sets of errors are uneven around mean value. Statistical characteristics of Ceres masses distributions induced idea to  compare  SM values with their corresponding MM values in the sense of correlation.
 We applied linear regression analysis to quantify the strength of the relationship (in the sense of correlation) between calculated Ceres masses from modified and standard method (our goal was not the prediction of results).
 We find that the results of SM and MM  satisfy  linear regression: $MM\sim (1.37\pm 0.61)+(0.69\pm0.13)SM$ (units are $10^{-10}M_{\odot}$) (Fig. 7), while their distribution express high correlation coefficient ($r=0.78$).
As it can be seen, obtained  values (based on modified method) are in good agreement with results obtained by standard method. We will explain in more details some of obtained results.

Close approach of Ceres and Pallas has been always interesting due to amount of masses of this two objects. However, unsuitable kinematical characteristics (significantly high relative velocity of the CE ) as well as unsuitable dynamical characteristics of Pallas orbit contributed to the variation in results obtained by many authors. Actually, the magnitude of the ballistic deflection is proportional to the mass of the perturbed body. Since Pallas is very large, it is not necessarily a good choice for mass determination.
 Values for Ceres mass based in this CE that we calculated are different about  $6\sigma$  from adopted Ceres mass ($4.76\,10^{-10} M_{\odot}$).
As it can be seen from Table 6, the observations from the period between 1825 and 1827 have been discarded (while other authors used the oldest observations). Standard method also produced improved value of Ceres mass due to large number of observations made after 1997 which are of high quality, while the oldest observations from 1825 were discarded based on  our criterion which we adopted for discarding (i.e. keeping) observations.

The case of (348) May is among the most gravitationally efficient (its amount is about $100^{\prime\prime}$ within the time interval covered by observations) CE  which took place in the Main asteroid belt. Applied methods produced results which satisfy $3
\sigma$ metrics relative to the adopted mass of Ceres, as well as relative to each other.

 CE with bodies numbered greater than 1000 usually occur at relative velocities nearby average relative velocity. Based on two methods, calculated individual Ceres masses are within  $3\sigma$ distance. The formal errors in these masses are considerably larger than for low numbered asteroids.
 Often, the uncertainty of a measurement is found by repeating the measurement enough times to get a good estimate of the standard deviation of the values. Then, any single value has an uncertainty equal to the standard deviation. However, if the values are averaged (in any kind of sense), then the mean measurement value has a much smaller uncertainty, equal to the standard error of the mean. In this context, uncertainty depends on both the accuracy and precision of the measurement.
 We will note that in our process of calculation Ceres mass used preencounter and postencounter observations had distribution of (O-C) residuals which are not significantly drawn from normally distributed population at $0.05$ level of confidence according to the Kolmogorov-Smirnov test.
 If corresponding orbits were preprocessed and smoothed according to the least square method (no significant systematic errors in observations)  and if the dynamical model is appropriate, then the formal error of the individual mass value should give good impression on actual uncertainty. Unfortunately,
 formal errors cannot account for systematic errors.
The systematic errors could originate from many sources, like the incompleteness
of the dynamical model, whose effect is shown by
Michalak (2000). This source is illustrated by deriving the masses
of three massive asteroids with or without including perturbing
asteroids in its dynamical model. The results show that the formal
errors of each determination remain quite similar in both
cases while the masses themselves change with the model.
In some cases, the correlation which can exist between the
a priori mass of certain perturbed asteroids and the perturber can
cause further error on the computed mass. \cite{Hi02} noted
the example of Pallas and stated that this correlation is due to the
similarity in the mean motions and mean longitude in the case
of Ceres perturbing Pallas. Goffin's analysis (2001) pointed out
other causes: the high inclination and eccentricity of Pallas is the
origin of the small number of significant close approaches for the
mass determination. However, we will emphasize here, that the problem remains and it is in practice very
difficult to estimate the value of the uncertainty in the derivation
of the masses of the asteroids by gravitational means.
Also, the statistical distribution of the relative precision of the masses is given in columns ${\sigma(m)/M}_{SM}$ and ${\sigma(m)/M}_{MM}$  in Table 6. It could be seen that all masses determined by MM have been estimated to better than $10\%$, there are 8 close encounters which produced precision better than $1\%$, and 19 close encounters provided precision better than $5\%$. In the case of SM, 9 values have relative precision better than $2\%$, and this number is rising to 19 for a $10\%$. However, in the case of SM all values are determined with relative precision better than $20\%$. The relative precision of weighted mean of  SM values are $1\%$, while for MM is $1.6\%$.
  Consequently, mentioned masses of highly numbered bodies we did not discarded because they do not differ more than $3\sigma$ from mean value, which means that calculated values  were not significantly influenced by systematic factors.
 The sets of preencounter observations are relatively small in comparison with other cases, while the lengths of corresponding time spans are about 1 year (with an exception of CE with (1847) Stobbe).

Having in mind all previous statements, one can say that the SM and MM masses are significantly different. Consequently, arises questions like which  is better and why.  Since the Dawn spacecraft recently released (\url{http://dawn.jpl.nasa.gov/mission/journal_08_11_11.asp}) a preliminary mass ($2.59\times10^{20}\, \rm{kg}$) for 4
Vesta, we could propose a follow-up experiment, applying these
same techniques to determine the mass of 4 Vesta, and evaluating which technique
better matches the Dawn value.
\begin{landscape}
\begin{table}
\centering
\caption{Calculated masses of  Ceres, based on standard (SM) and modified method (MM).
Statistical characteristics of used sets of observations. In columns: $N_{1}$, $N_{2}$, $N_{3}$, $N_{4}$, $N_{5}$
are given respectively: total amount of observations, the number of discarded preencounter observations, the number of used preencounter observations, the number of discarded postencounter observations, the number of used post encounter observations. $T_{1}$ and $T_{2}$ the length of preencounter and postencounter time spans covered by observations.${\sigma(m)/M}_{SM}$ and ${\sigma(m)/M}_{MM}$ are relative precision of masses determined by SM and MM respectively.  $RMS_{1}$ is rms of preenncounter orbit and $RMS_{2}$ is rms of postencounter orbit.$|\Delta\alpha|$ and $|\Delta \delta|$ are largest absolute value of differences in geocentric rectascension and declination.}

\label{cm}
\centering
\resizebox{23cm}{!} {
\begin{tabular}{llllllllllllllll}

\hline
       Perturbed& $ N_{1}$ & $N_{2}$&$N_{3}$ &$T_{1}$&$RMS_{1}$            &$N_{4}$&$N_{5}$& $T_{2}$&$RMS_{2}$&SM& ${\sigma(m)/M}_{SM}$&MM& ${\sigma(m)/M}_{MM}$&$|\Delta\alpha|$&$|\Delta \delta|$\\

       asteroid&           &         &       &       &  [$^{\prime\prime}$]&       &       &       &[$^{\prime\prime}$]&[$ 10^{-10}\, M_{\odot}$]&$\%$& [$10^{-10}\,M_{\odot}$]& $\%$&[$^{\prime\prime}$] & [$^{\prime\prime}$]\\

            \hline

   (2) Pallas  & 7428& 177 &503 & 1827-1929&  0.89 &1045&5703 &1940-2002 & 0.42 &$4.45\pm0.05$&  1.12 &$4.22\pm0.04$& 0.95 &28&7.5\\

  (32) Pomona  &598& 129 & 162 & 1864-1974& 1.24    &35     &272  &1977-2002&  0.61 &$5.32\pm0.16$&3.01 &$5.18\pm0.05$& 0.97&30&10 \\

  (76) Freia    & 1012& 126 & 151 &1864-1974&  0.81  &  94  &641  & 1958-2002& 0.59  &$4.27\pm0.08$& 1.87 &$4.14\pm0.06$&1.45  &26&7.5\\

  (91) Aegina  & 716& 165  & 158 & 1866-1972& 0.90  &  53 &340  &1974-2002&  0.46 & $4.91\pm0.04$&0.81 &$5.00\pm0.02$& 0.4&80&22\\

  (203) Pompeia& 512& 61 & 58& 1879-1947& 2.09   &58  &335  &1949-2002& 0.61  &$4.73\pm0.04$& 0.85 &$4.79\pm0.02$&0.42 &96&35 \\

   (348) May   & 601& 56 & 65&1892-1983& 1.59   &58  &422  &1989-2002& 0.37   &$4.74\pm0.05$& 1.05 &$4.77\pm0.01$&0.21 &110&38 \\

   (347) Pariana &435& 35 & 35&1892-1937&  1.15  &69  &296  &1947-2002& 0.46  &$4.80\pm0.09$& 1.88 &$4.72\pm0.05$&1.06 &30&15  \\

   (454) Mathesis& 646& 112 & 101&1900-1970&1.39   &50  &383  & 1974-2002& 0.50 &$4.48\pm0.06$& 1.40 &$4.33\pm0.01$&0.23 &60&30 \\

    (488) Kreussa& 520& 76 & 52&1901-1961& 1.30  &57  &335  &1965-2002& 0.40  &$4.64\pm0.16$& 3.45 &$4.26\pm0.11$& 2.58&20&6  \\

    (534) Nassovia& 563& 91 & 54&1904-1974& 1.34  &56  &362  &1983-2002& 0.46  &$4.83\pm0.07$& 1.45 &$5.12\pm0.04$&0.78 &55&20 \\

(548) Kressida& 465& 46 & 59&1909-1980&  2.30  &50  &310  &1983-2002& 0.44  &$5.28\pm0.24$& 4.55 &$4.89\pm0.10$&2.04 &17.5&5 \\

(621) Werdandi& 542& 13 &20 & 1911-1958 & 1.56  &36  &473  & 1966-2002&0.62   &$4.35\pm0.15$& 3.45 &$4.56\pm0.20$& 4.39&22.5&6 \\

(792) Metcalfia& 413& 10 & 38&1915-1947&19.74   &38  &327  &1953-2002&  0.67 &$5.81\pm1.10$& 18.93 &$5.22\pm0.35$&6.70 &20&7.5  \\

  (850) Altona     & 342& 29 & 22&1917-1964& 2.40  &34  &257  &1972-2002&0.60   &$4.91\pm0.16$& 3.26 &$4.68\pm0.11$& 2.35&32.5&7 \\

   (1642) Hill    & 401& 2 & 3& 1908&  - &67  &329  &1931-2002& 0.63  &$4.81\pm0.06$& 1.25 &$4.81\pm0.08$&1.66 &50&20 \\

  (1847) Stobbe   &365& 13 & 19&1902-1952&  2.25 &42  &291  &1973-2002&  0.47 &$3.94\pm0.23$& 5.84 &$4.10\pm0.17$&4.15 &18&7  \\

   (3344) Modena  & 204& 1 & 1& 1955& -   &  13  &189  &1982-2002&0.80  &$4.34\pm0.38$& 8.76 &$4.36\pm0.11$&2.52 &16&5  \\

   (2051) Chang &458 &4  &1 &1933 &  - &51 &402  &1955-2002&  0.63 &$4.63\pm0.26$ & 5.62 &$4.57\pm0.21$& 4.60&10&3 \\

  (6010) Lyzenga  &173& 1 & 1&1953& -  &5 &166  &1974-2002 & 0.53   &$4.60\pm0.29$& 6.30 &$4.48\pm0.16$&3.57 &17.5&6 \\

  (6594) Tasman  & 218&1  &1  &  1954& -  &15 &201  &1987-2002 & 0.55  &$5.02\pm0.44$&8.76 &$5.22\pm0.49$&9.39 &10&2.5 \\

  (34755) 2001QW120 &100&2 &1  &  1950 & - & 6  &91  &1985-2001 & 0.72   & $5.05\pm0.76$& 15.05 &$4.31\pm0.01$&0.23 &17.5&4\\

          \hline
     \end{tabular}
  }
  \end{table}
\end{landscape}

\begin{figure*}
\centering
\includegraphics[width=8.8cm]{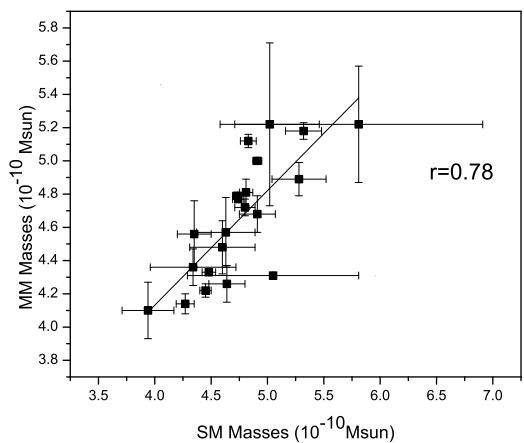}
\caption{Distribution of SM and MM values of Ceres mass. Correlation coefficient r is shown on right part of the plot. Solid line is fitted linear regression  with intercept:$1.37\pm0.61$, slope:$0.69\pm0.13$).}
\label{ceres2}
\end{figure*}



\begin{figure*}
   \includegraphics[width=8.8cm]{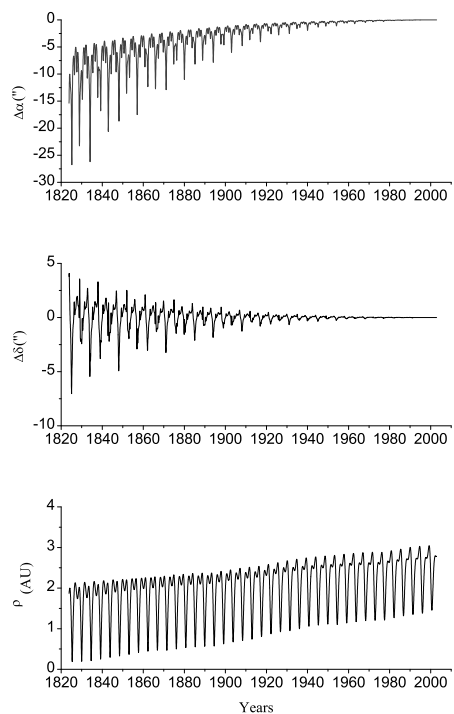}
 \caption{From top to bottom: Perturbations in right ascension and declination caused by Ceres on Pallas, dynamical evolution of mutual distance between Ceres and Pallas (bottom). }
\label{cerespalas}
\end{figure*}

\begin{figure*}
\includegraphics[totalheight=5.3cm]{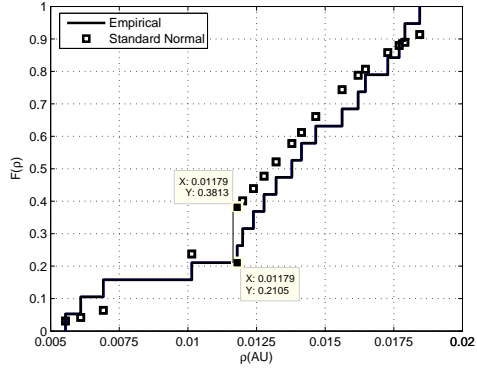}

\caption{Kolmogorov-Smirnov test applied on set of mutual distances ($\rho$) of future CE with Ceres: A sample distribution function (solid line) is compared to a given probability distribution function (squares). The test statistic  is the maximum difference between the curves. This maximum of 0.17 occurs as the data approaches $x=0.001179$ from below. The empirical curve has the value 0.2105 here.} \label{KS0}
\end{figure*}
\begin{figure*}
\includegraphics[totalheight=5.3cm]{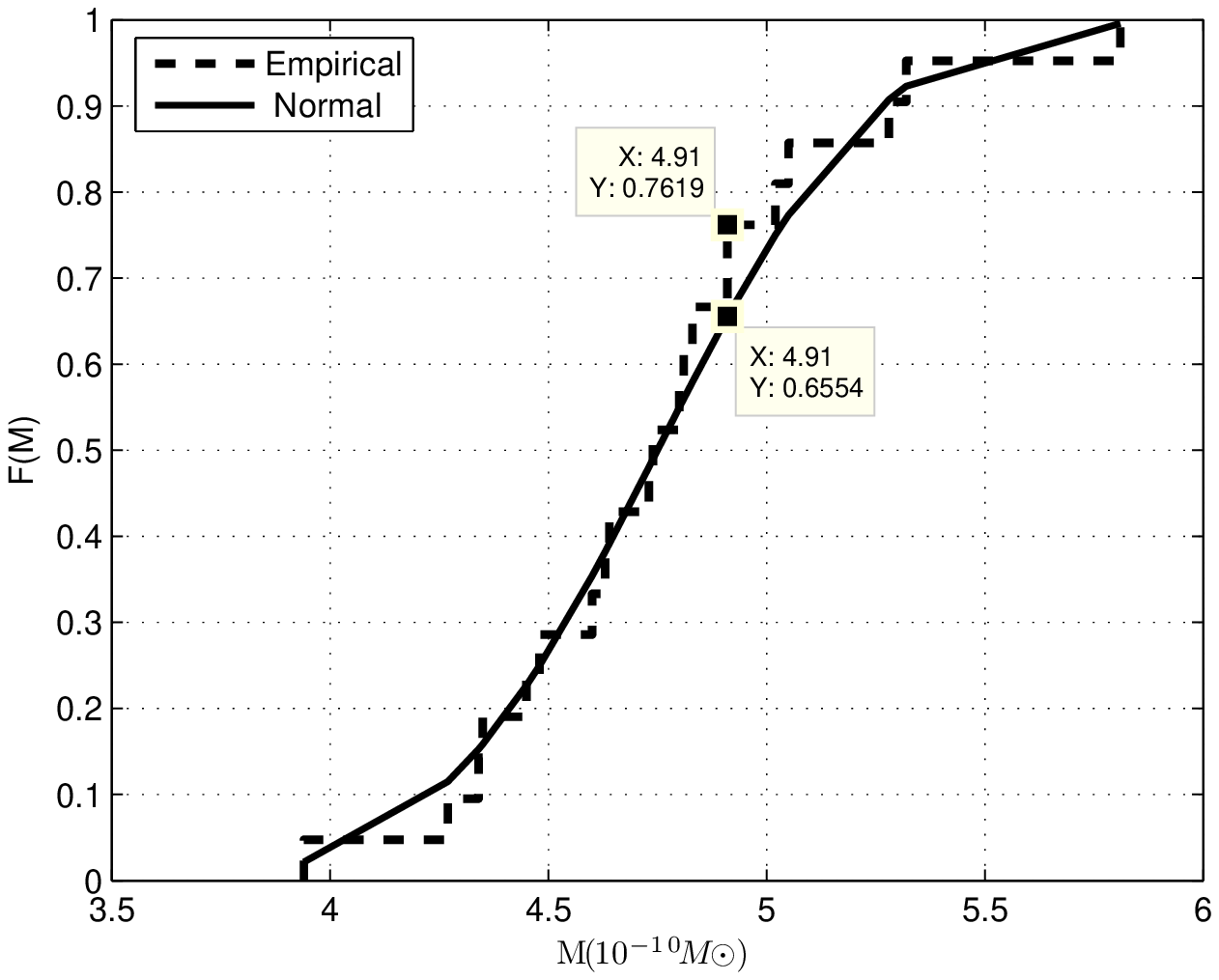}
\includegraphics[totalheight=5.3cm, origin=c]{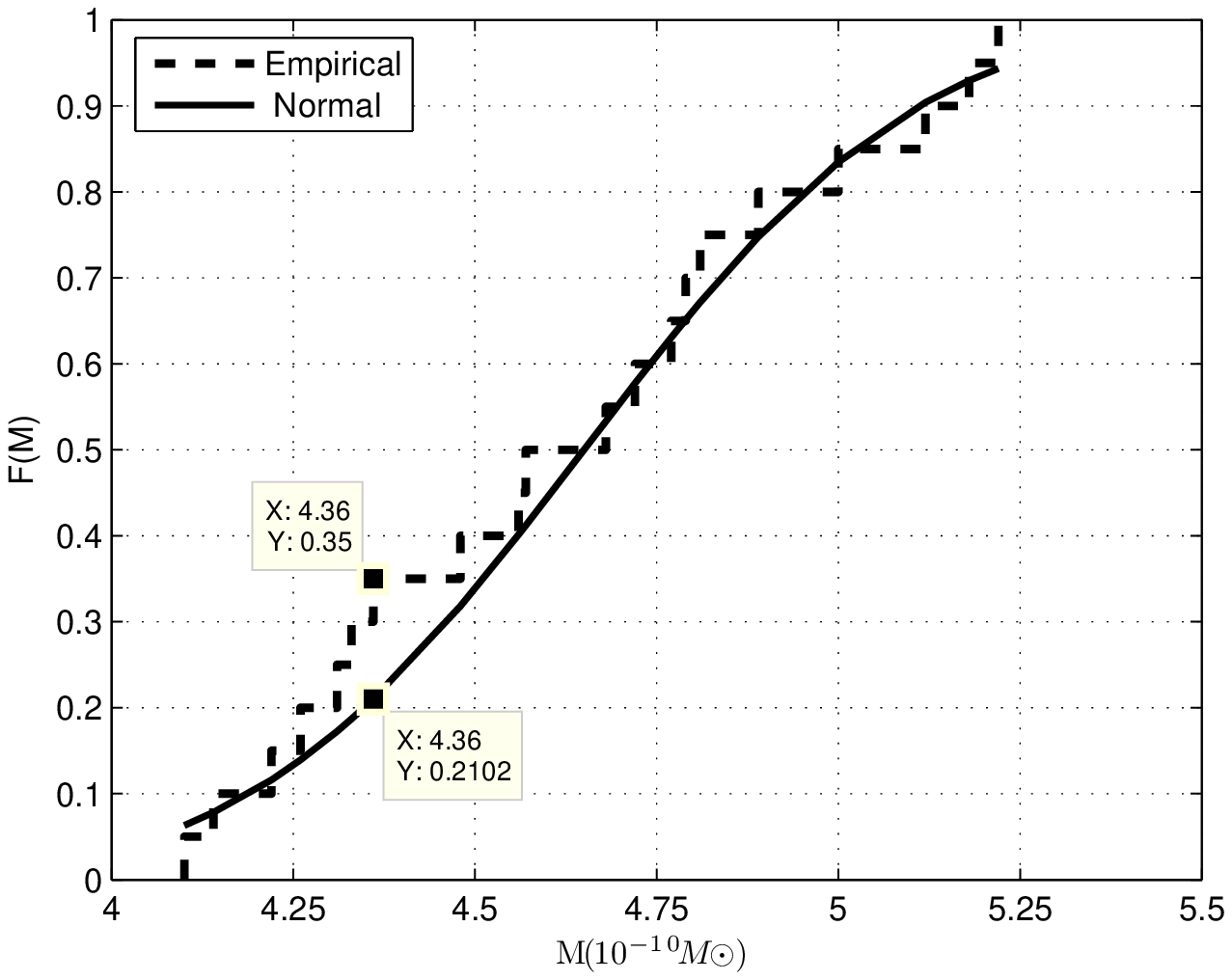}
\caption{Kolmogorov-Smirnov test: A sample distribution function (dashed line
line) is compared to a given probability distribution function (solid line). The test statistic is the maximum difference between the curves. Points where maximum occurred  are labeled.Left panel: KS test applied on set of SM Ceres masses.
Right panel:KS test applied on set of MM Ceres masses. } \label{KS}
\end{figure*}
\section{Conclusion}

We made a list of asteroids useful for determination of Ceres mass from mutual perturbations. In the list are included both: historically well known perturbed bodies and several asteroids found by our searching procedure.
Then the Ceres masses were calculated  independently for all test asteroids by means of standard method and our modified method.
Results were examined by means of smallest formal errors and rejecting those which do not satisfy $3\sigma$ metric (i.e. influenced by systematic factors).

The resulting mass of Ceres based on standard method is $(4.70\pm0.04)\,10^{-10}M_{\odot}$ which is within  standard $3\sigma$ distance from adopted value of Ceres mass. Weighted mean value of Ceres mass, obtained from modified method, is $(4.54\pm0.07)\,10^{-10}M_{\odot}$.
We found that the results obtained by two methods are highly correlated. Based on this, one can say that the new modified method, which is not tested widely so far, could be successfully used for asteroid mass determination.

Generally, the masses we found appears to be smaller  by about several percent than the  value recommended by the IAU.

\section*{Acknowledgments}

The author is grateful for the support provided by the Ministry of Education and Science of the Republic of Serbia through the project
'Influence of collisional processes on astrophysical plasma spectra' (176002) and for the help of prof. dr Milan S. Dimitrijevi{\'c}.

\end{document}